\documentclass{svjour3}
\smartqed   
\usepackage{graphicx}

\journalname{Few-Body Systems}
\begin{document}

\title{Phase transition in the SRG flow of nuclear interactions
\footnote{Presented by VST at the 23rd European Conference on Few-Body Problems in Physics, Aarhus, Denmark, 08 - 12 August 2016.} }

\author{V. S. Tim\'oteo         \and
        E. Ruiz Arriola \and  S. Szpigel
}

\institute{V. S. Tim\'oteo \at
              Faculdade de Tecnologia, Universidade Estadual de Campinas - UNICAMP \\
             13484-332, Limeira, S\~ao Paulo, Brasil
           \and
              E. Ruiz Arriola \at
              Departamento de Fisica At\'omica, Molecular y Nuclear and Instituto Carlos I de Fisica Te\'orica y Computacional, Universidad de Granada \\
E-18071  Granada, Spain            
           \and
           S. Szpigel \at
           Centro de R\'adio-Astronomia e Astrof\'isica, Universidade Presbiteriana Mackenzie \\
           01302-907, S\~ao Paulo, SP, Brasil
}
\date{Received: date / Accepted: date}

\maketitle

\begin{abstract}

We use a chiral interaction at N3LO in the $^1S_0$ channel of the
nucleon-nucleon interaction in order to investigate the on-shell
transition along the similarity renormalization group flow towards the
infrared limit. We find a crossover at a scale that depends on the
number of grid points used to discretise the momentum space.


\end{abstract}

One of the most appealing features of nature is universality. Some
phenomena disguise themselves across many different areas where
physical systems are described by sometimes unrelated theories or
models. Yet they appear recurrently in some form.

An excellent example is the phase transition resulting from a broken
symmetry.  It is observed in magnetism when the temperature of a spin
chain in a two-dimensional Ising model crosses a critical value
\cite{ONSA}. It appears in nuclear physics when observing rotational
spectra of deformed nucleiIt \cite{FRAU} and it is also present in
hadron physics when the coupling between quarks in a two-flavour NJL
model exceeds a critical value \cite{NJL}.

In both magnetism and nuclear physics the phase transition results
from the breaking of the rotational symmetry and the corresponding
Goldstone bosons are spin waves and nuclear rotation. In hadron
physics the phase transition results from the chiral symmetry breaking
and the corresponding Goldstone boson is the pion. This phenomenon is
shown in Fig. \ref{trans} for both two-dimensional Ising model (left)
and two-flavour NJL model (right).

In this work we report on a similar and remarkable phase transition
observed in the similarity renormalization group flow, which is used
to change and calibrate the resolution scale of nuclear interactions
to their natural values in different applications.

\begin{figure}
\includegraphics[width=6cm]{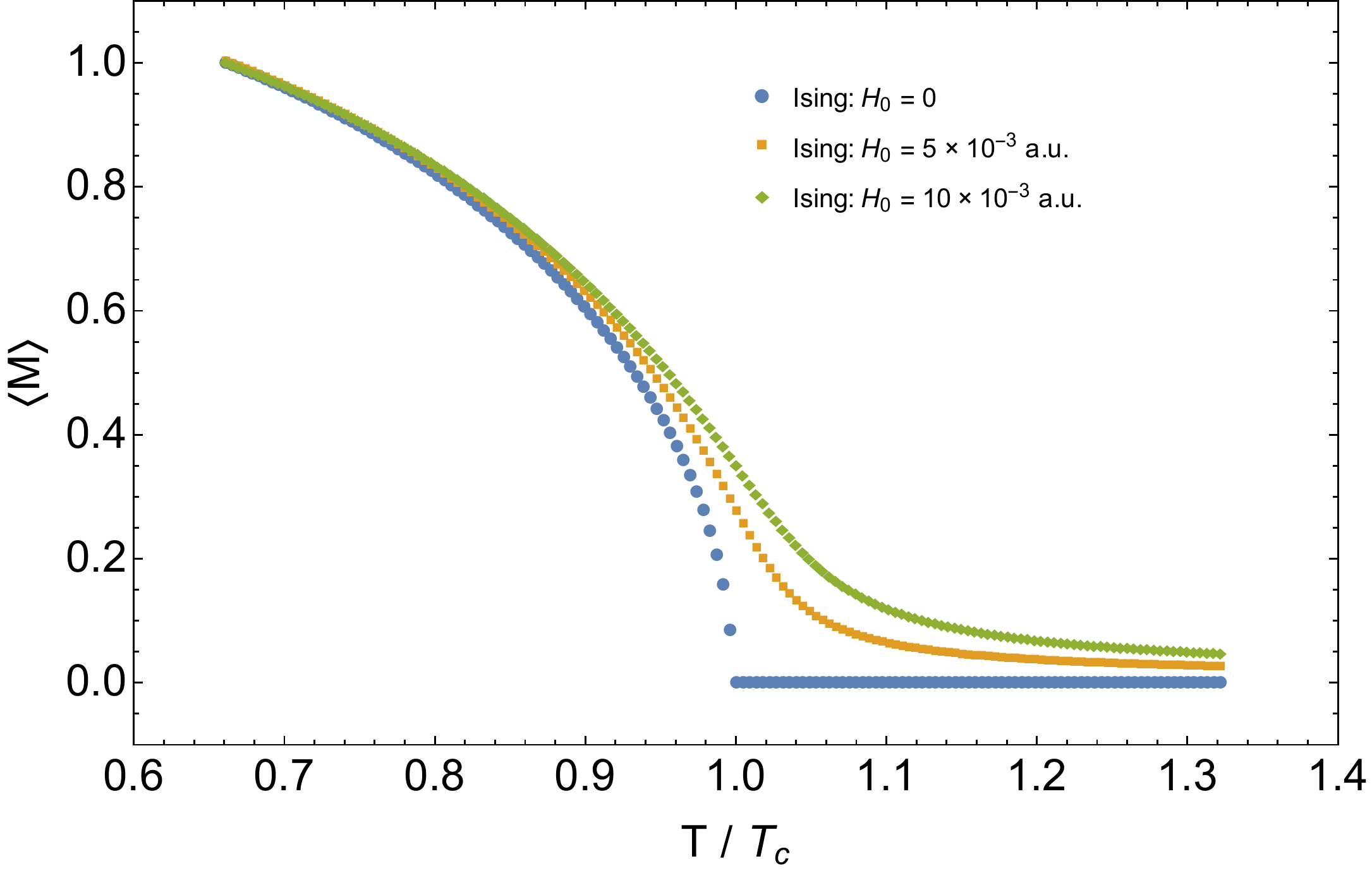} 
\includegraphics[width=6cm]{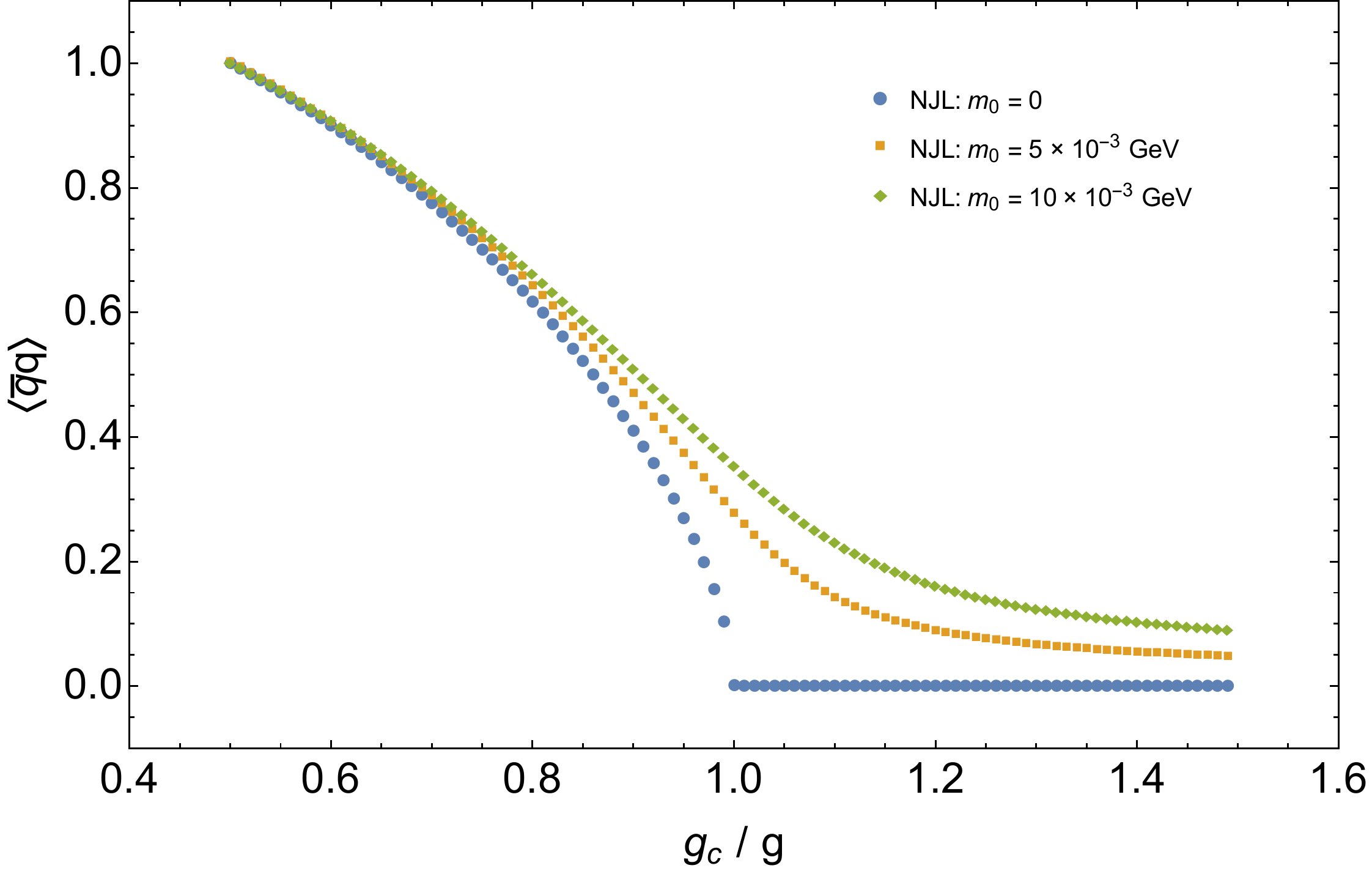} 
\caption{Universal phase transitions: the Ising model (left) and the NJL model (right). 
The critical value separates the Wigner-Weyl and Nambu-Goldstone phases. 
Order parameters are normalized so that they start from one.}
\label{trans}
\end{figure}

The evolution of an NN interaction with the SRG \cite{BFS} is
performed by numerically integrating the Wegner renormalization group
flow equation for the potential matrix
\begin{eqnarray}
\frac{d~V_s(p,p')}{ds} &=& - (p^2-p'^2)^2~V_s(p,p') +\frac{2}{\pi} \int_0^\infty dq~q^2\nonumber \\
 &\times& (p^2 + p'^2 - 2 q^2)~V_s(p,q)~V_s(q,p')
\end{eqnarray}
where $s=1/\lambda^4$ and $\lambda$ is the similarity cutoff. The flow
equation generates a set of isospectral interactions that approaches a
diagonal form as $s\to\infty$ (or $\lambda\to0$).  Here we explore the
use of the Frobenius Norm (FN) of the potential as a measure of the
on-shellness of the interaction in order to study the on-shell
transition as the interaction undergoes the renormalization group flow
through Wegner’s equation. For a real self-adjoint potential,
$V_{\lambda}(p',p)=V_{\lambda}(p,p') $ the FN, $\phi_\lambda$, is
defined by means of the formula
\begin{eqnarray}
\phi_\lambda^2 = || V_\lambda  ||^2  = \left(\frac{2}{\pi} \right)^2 \int_0^\infty dq~q^2\int_0^\infty dp~p^2 \left[ V_\lambda (p,q) \right]^2.
\end{eqnarray}
As an order parameter for the on-shell transition, we consider the
derivative of the Frobenius norm with respect to the similarity cutoff
\begin{equation}
\beta = \frac{\partial\phi}{\partial\lambda} \; ,
\end{equation}
and to find the transition scale we look at the derivative of the
order parameter
\begin{equation}
\eta = \frac{\partial\beta}{\partial\lambda} \; .
\end{equation}
This procedure is same as the one used to study, e.g., the chiral
(crossover) transition of QCD, where the order parameter is the chiral
condensate, the running scale is the temperature and the transition
temperature is obtained from the thermal susceptibility, which is the
derivative of the condensates with respect to the temperature
\cite{FTAPK}.

\begin{figure}
\begin{center}
\includegraphics[width=4cm]{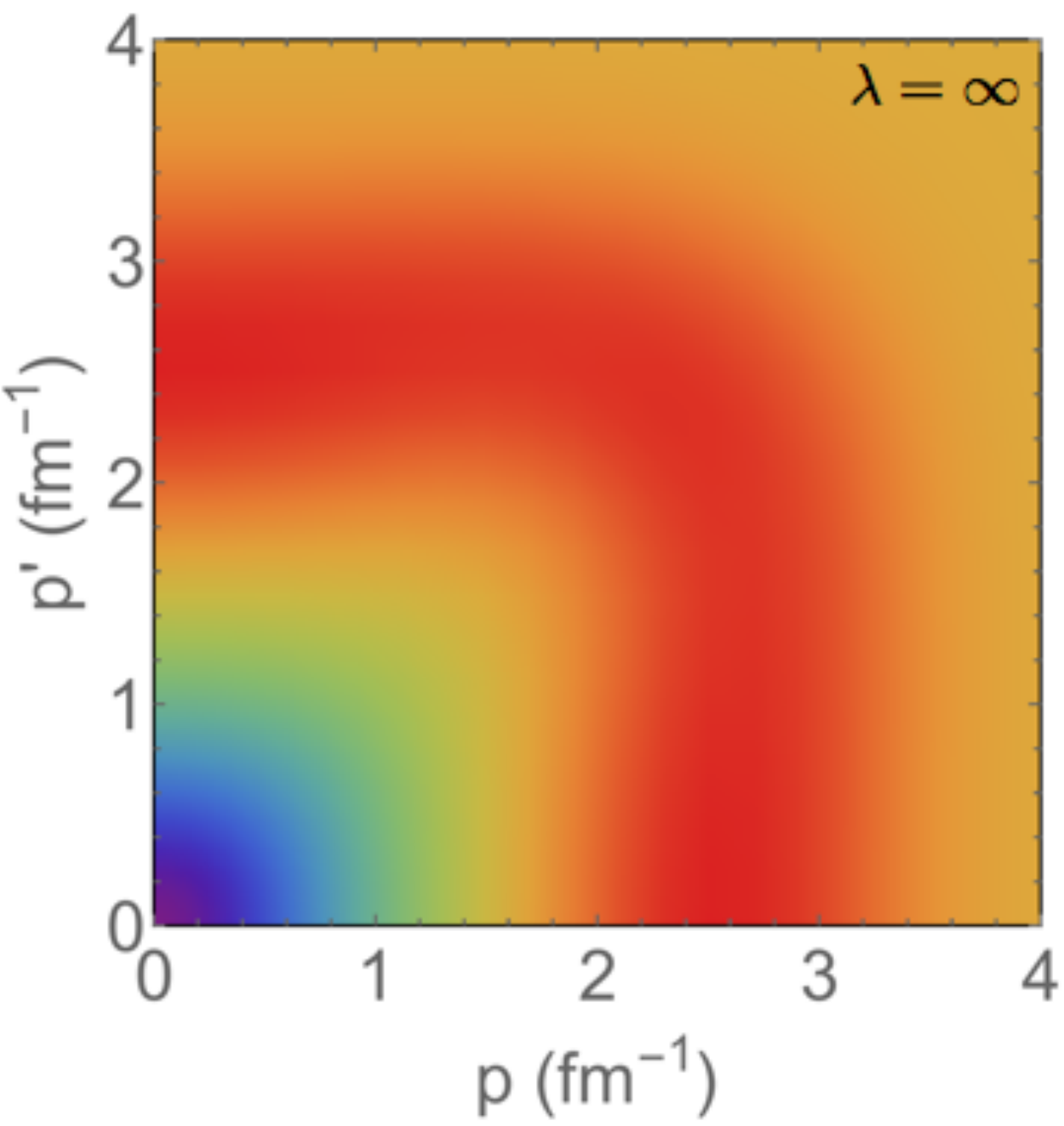} \hspace*{1cm}
\includegraphics[width=4cm]{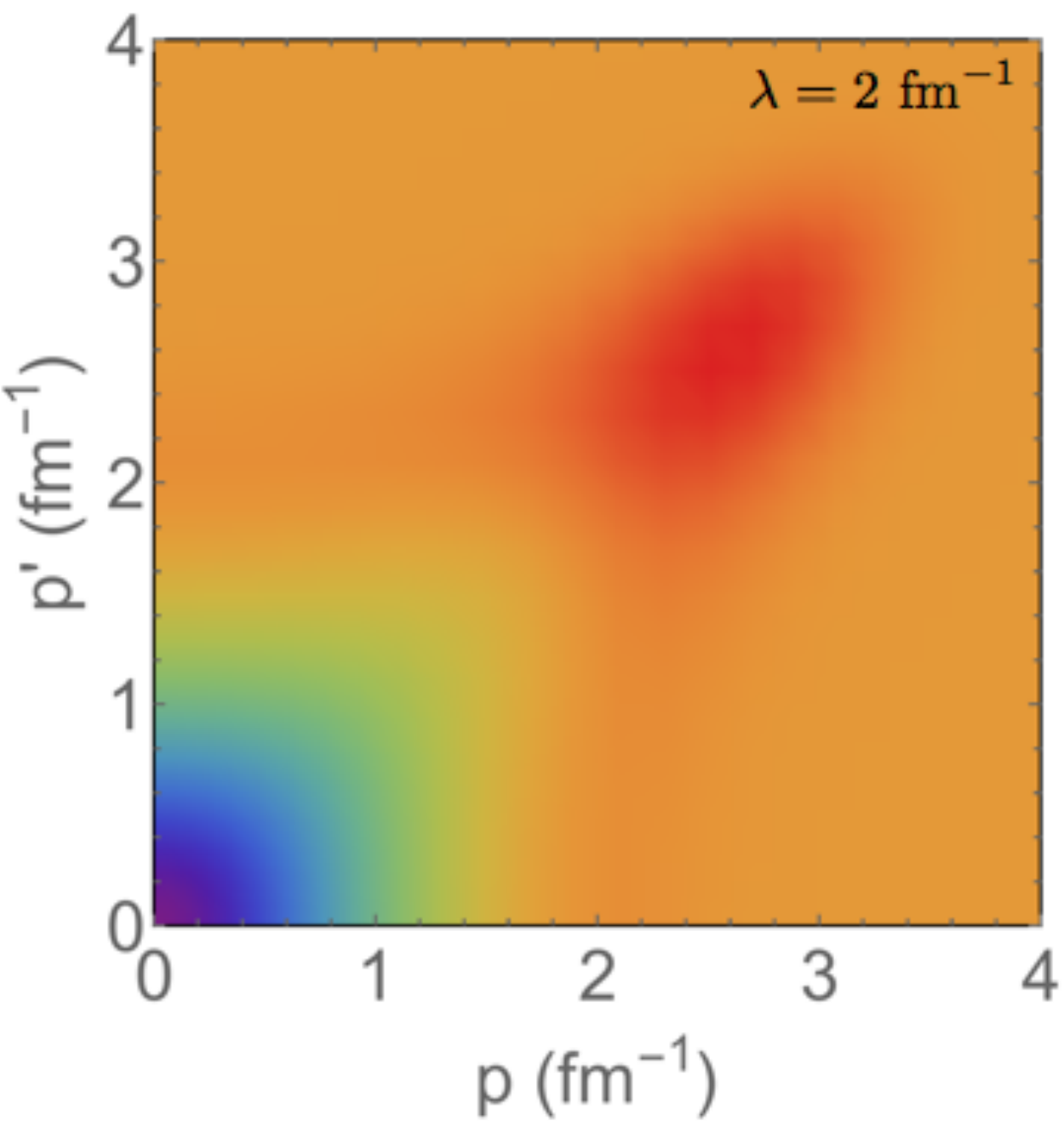}  \\ \vspace*{0.5cm}
\includegraphics[width=4cm]{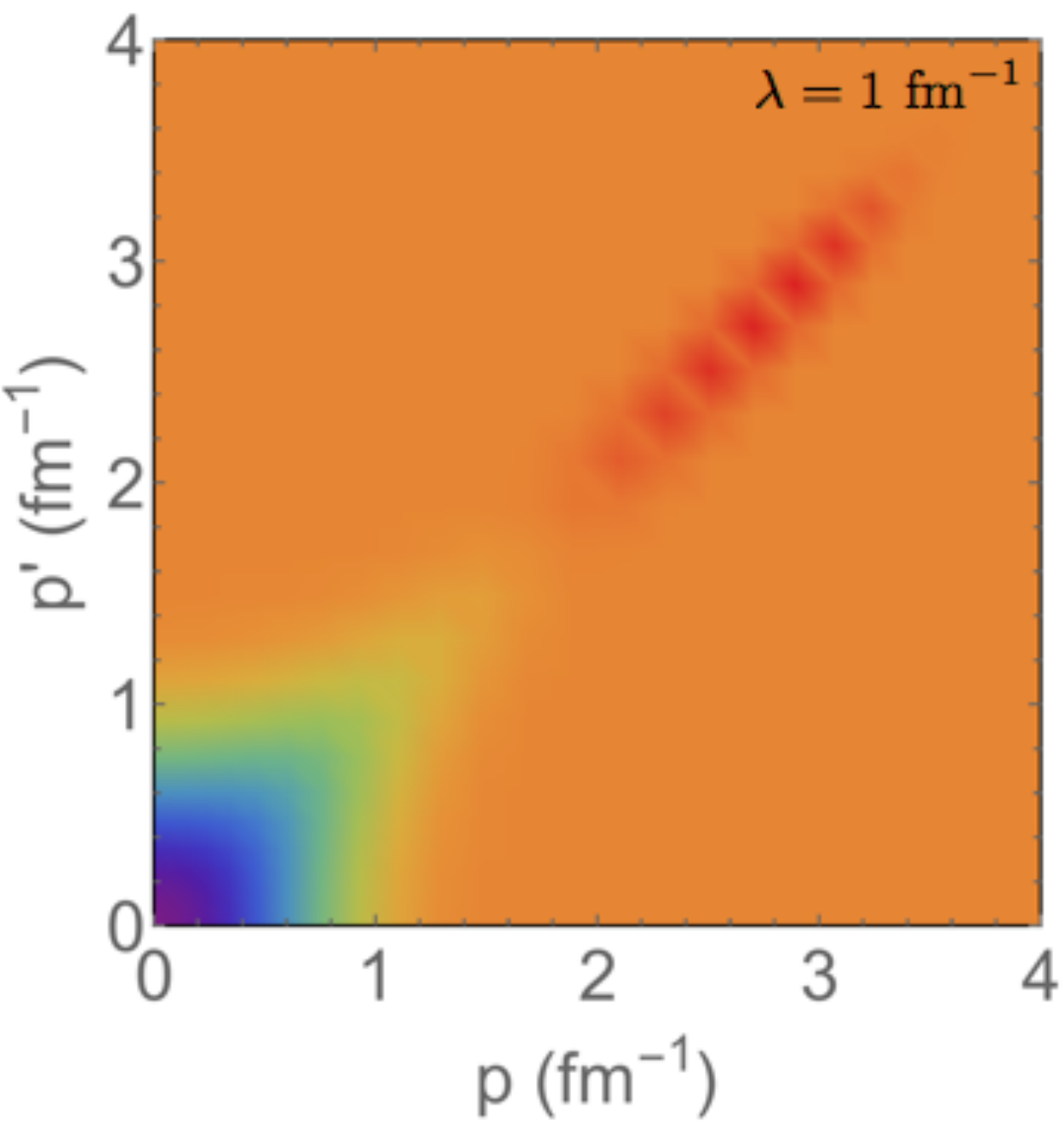} \hspace*{1cm}
\includegraphics[width=4cm]{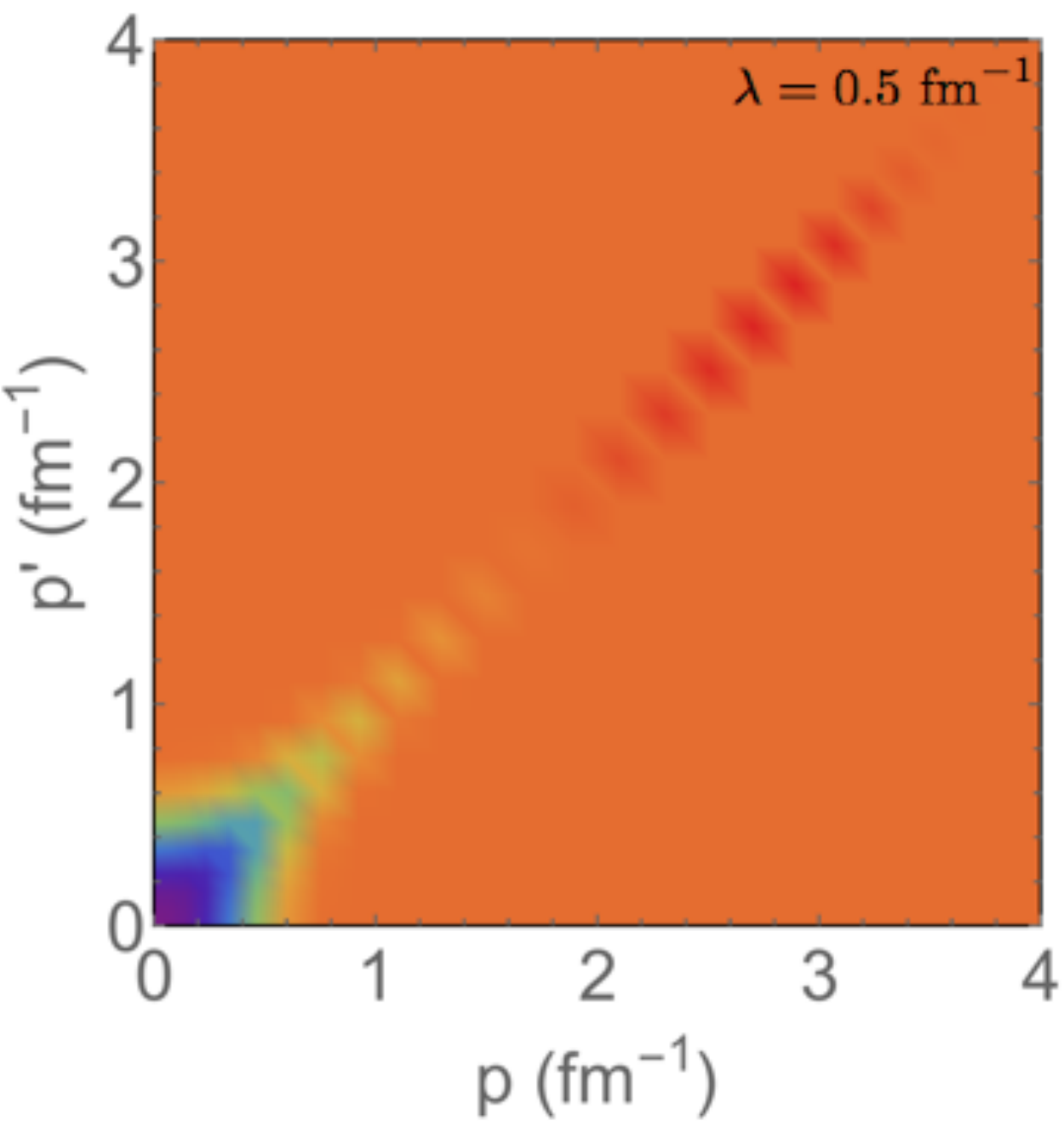} 
\end{center}
\caption{SRG evolution of the N3LO chiral potential in the $^1S_0$ 
channel towards the infrared region of the similarity cutoff.}
\label{evo}
\end{figure}

Since we are interested in the infrared region of the similarity
cutoff, we choose the N3LO chiral potential of Entem and Machleidt
\cite{ME} due to its short tail in momentum space which makes the SRG
evolution for $\lambda\to0$ more practical. In the case of the N3LO
potential, we can work with a maximum momentum of $4~{\rm fm}^{-1}$
and $N=10,20,30$ points for discretisation. This setup gives very
reasonable values of the two-body contribution for the Triton and
Helium binding energies as we approach typical values for the
similarity cutoff. Following our previous works
\cite{AST1,AST2,AST3,AST4}, where a toy model for the nuclear force
was used, we perform here the evolution of the N3LO potential in the
range $0<\lambda<2~{\rm fm}^{-1}$, generating a family of
phase-equivalent effective potentials $V_\lambda (p,p')$ partially
shown in Fig. \ref{evo}.

In the upper left panel of Fig. \ref{trans} we show the running of the
Frobenius norm with the similarity cutoff for different numbers of
grid points. The norm drops very rapidly till it becomes stationary,
indicating the onset of the on-shell limit. The order parameter
$\beta$ is displayed in the upper right panel of Fig. \ref{trans},
where we observe a clear crossover. We can locate the transition scale
by looking at the derivative of the order parameter. We call this
quantity “similarity susceptibility” and denote it by $\eta_\lambda$,
in analogy to the thermal susceptibility $\chi_T$ of QCD. The
similarity susceptibility is shown in the lower left panel of
Fig. \ref{trans}, revealing the expected behaviour for a crossover
transition: a well-defined peak of the susceptibility at a similarity
cutoff $\lambda_c$. As the number of grid points increases, the peaks
of the similarity susceptibility become narrower, higher and at lower
$\lambda_c$. For N = 30 grid points, the peak of $\eta_\lambda$ is at
$\lambda_c \sim 0.9 ~ {\rm fm}^{-1}$. The critical similarity
susceptibility for different number of grid points is displayed in the
lower right panel of Fig. \ref{trans}.
\begin{figure}[h]
\begin{center}
\includegraphics[width=4cm]{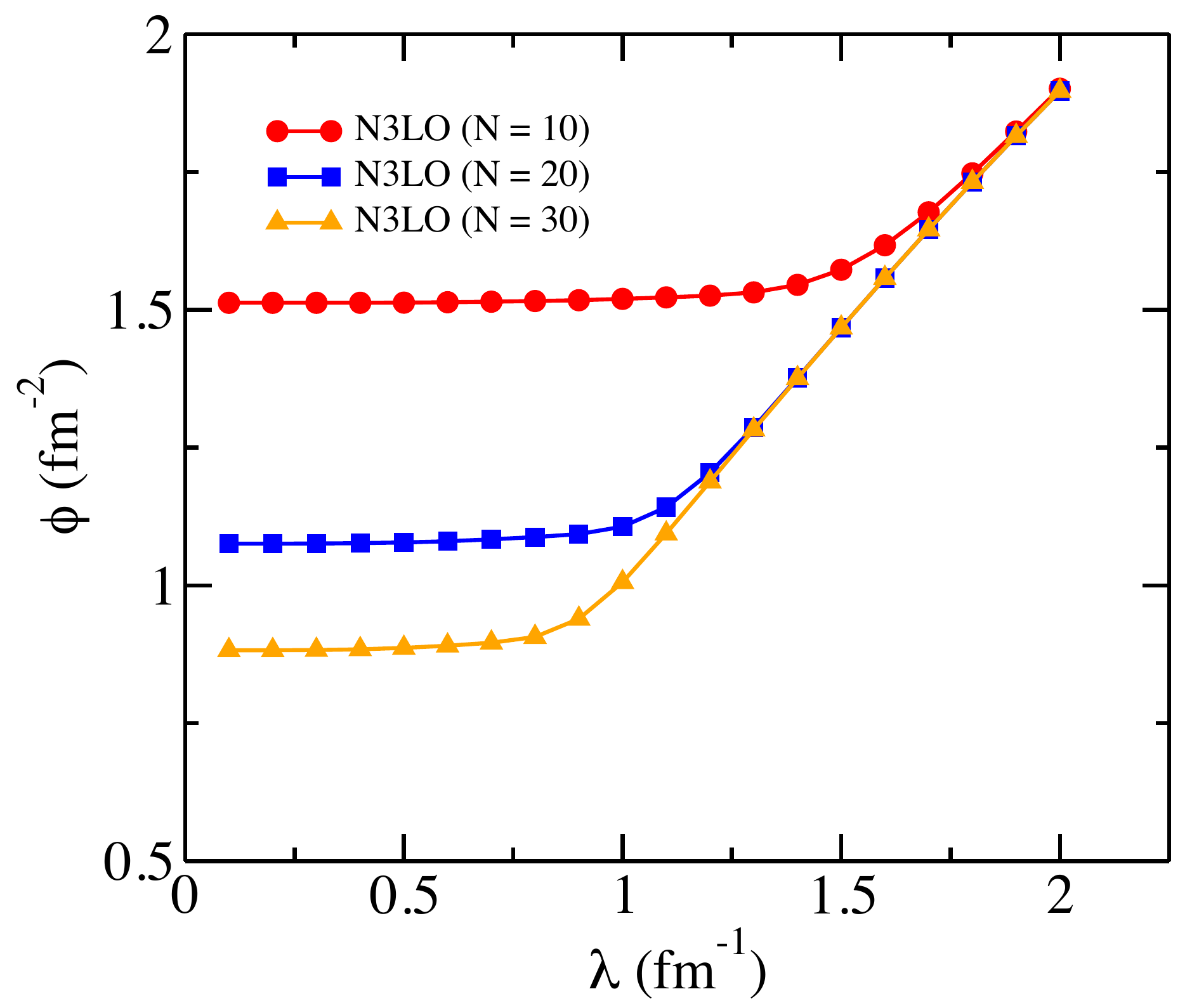} \hspace*{0.7cm}
\includegraphics[width=4cm]{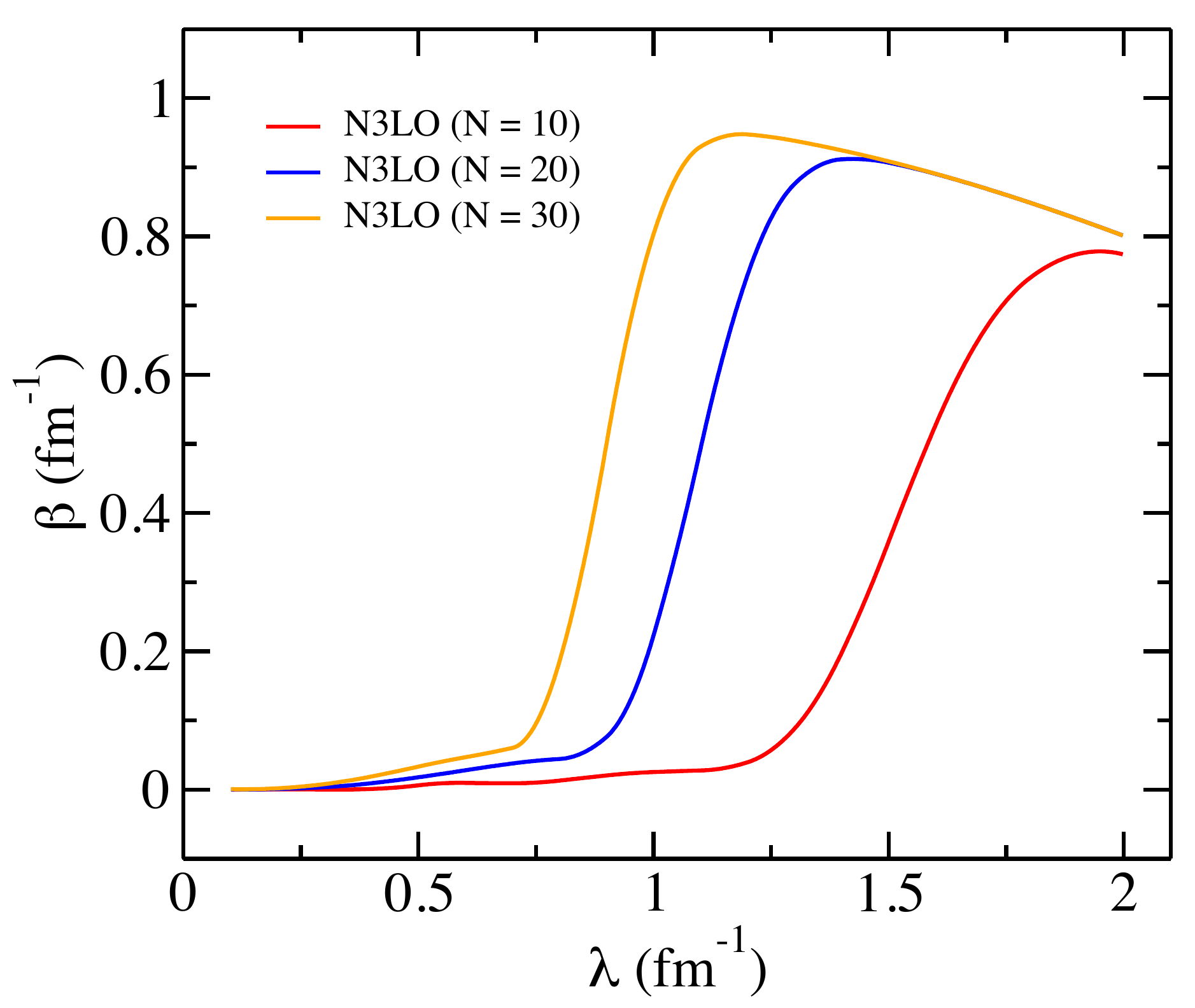} \\ \vspace*{0.7cm}
\includegraphics[width=4cm]{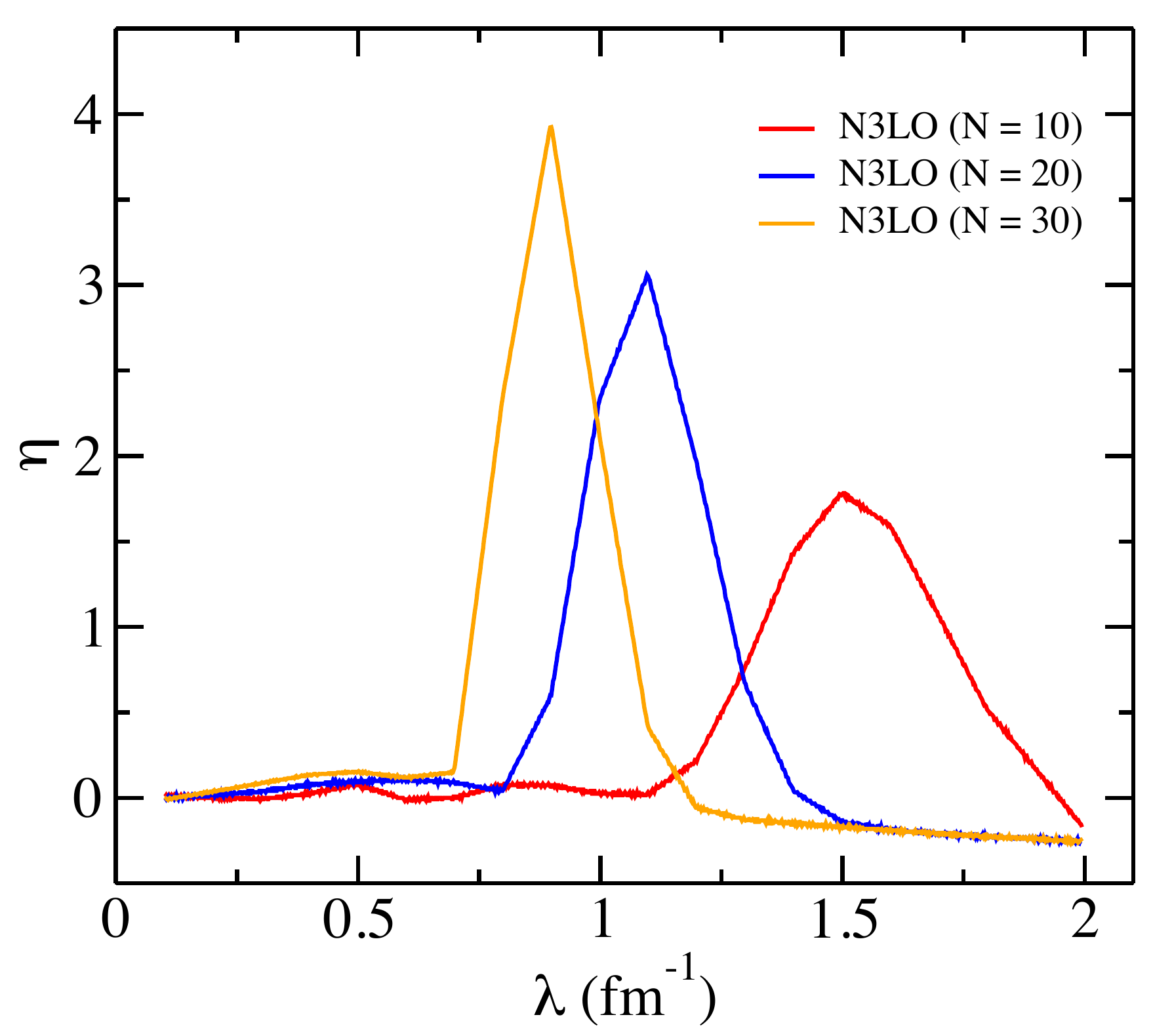}   \hspace*{0.7cm}
\includegraphics[width=4cm]{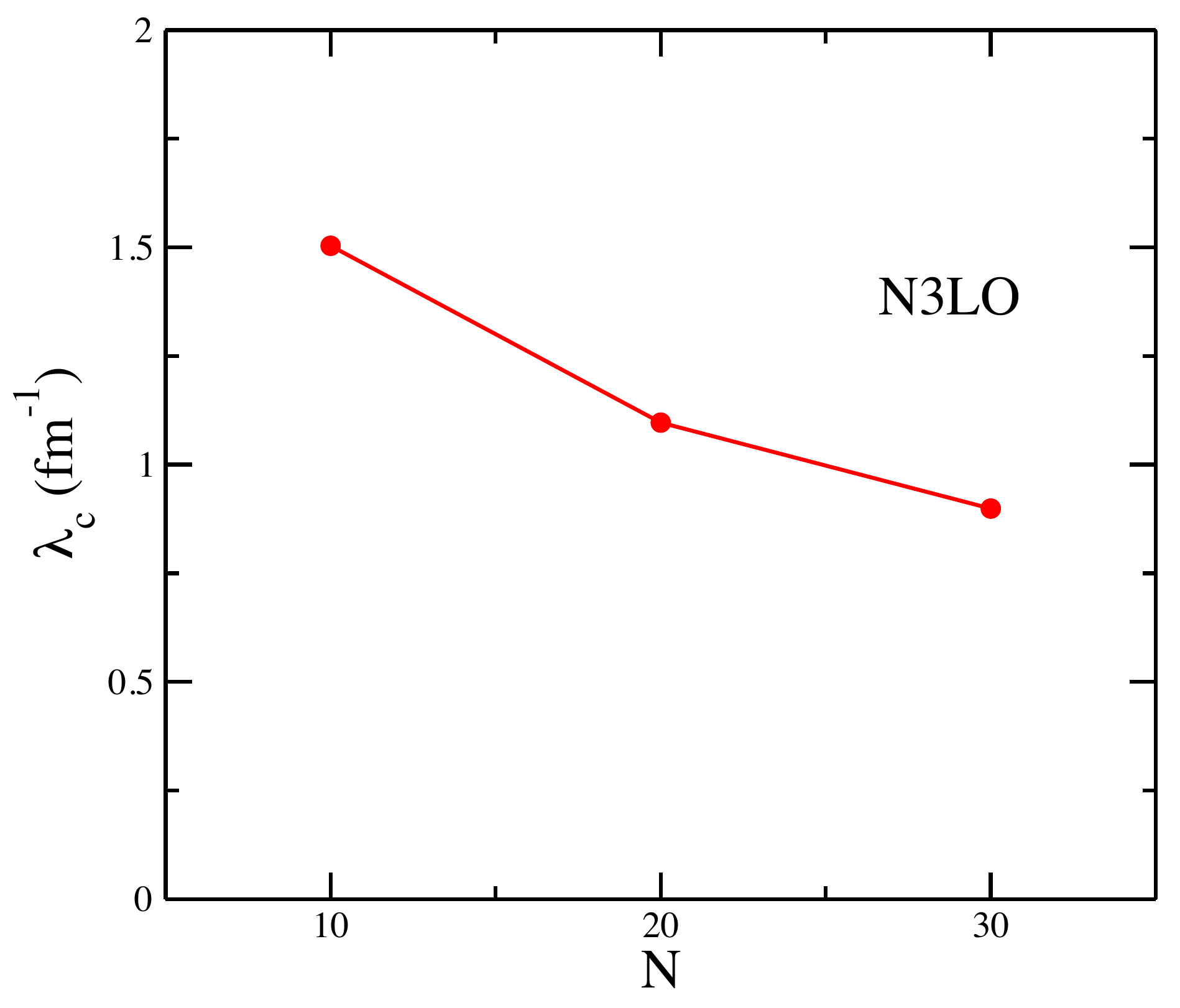} 
\end{center}
\caption{Frobenius norm $\phi$ (upper left), order parameter
  $\beta=\frac{\partial\phi}{\partial\lambda}$ (upper right) and
  similarity susceptibility
  $\eta=\frac{\partial\beta}{\partial\lambda}$ (lower left) as
  functions of the similarity cutoff and the critical similarity
  cutoff $\lambda_c$ as a function of the number of grid points $N$
  (lower right).}
\label{trans}
\end{figure}

Summarising, we have studied the on-shell transition of the similarity
renormalization group flow for a N3LO chiral potential. Our results
indicate a crossover at a critical similarity cutoff $\lambda_c$
approximately $0.9~{\rm fm}^{-1}$, below which we can consider the
interaction to be on-shell. The usefulness of this observation is that
in a system with a finite size, such as the atomic nucleus, where the
momentum is naturally discretized, the on-shell limit is reached
rather quickly.  The finite momentum resolution $\Delta p$ imposes an
infrared cut-off which effectively builds the on-shell limit for
$\lambda \gg \Delta p$.

\section*{Acknowledgements}

We would like to thank Spanish Mineco (FIS2014-59386-P),
Junta de Andalucia (FQM225), FAPESP (2016/07061-3 \& 2016/05554-2) and 
CNPq (306195/2015-1).

\end{document}